\documentclass{ws-ijqi}

\usepackage{amsfonts,amssymb}
\usepackage{overcite}

\expandafter\let\csname url\endcsname\relax
\usepackage{url}
\expandafter\let\csname trimmarks\endcsname\relax

\newcommand{\<}{\langle}
\renewcommand{\>}{\rangle}

\newcommand{\be}{\begin{equation}}
\newcommand{\ee}{\end{equation}}
\def\ba#1\ea{\begin{align}#1\end{align}}
\newcommand{\nn}{\nonumber\\}

\newcommand{\ra}{\rightarrow}

\newcommand{\cra}{\Rightarrow}
\newcommand{\cla}{\Leftarrow}

\newcommand{\cP}{{\cal P}}
\newcommand{\cA}{{\cal A}}
\newcommand{\cB}{{\cal B}}
\newcommand{\E}{{\cal E}}
\newcommand{\M}{{\cal M}}
\newcommand{\N}{{\cal N}}
\renewcommand{\r}{\rho}
\newcommand{\s}{\sigma}
\def\e{{\epsilon}}
\newcommand{\D}{\Delta}
\newcommand{\ot}{\otimes}

\def\bbR{{\mathbb R}}
\def\bbN{{\mathbb N}}

\newcommand{\eq}[1]{Eq.~(\ref{eq:#1})}

\newcommand{\eqssss}[4]{Eqs.~(\ref{eq:#1}), (\ref{eq:#2}),
  (\ref{eq:#3}), and (\ref{eq:#4})}
\renewcommand{\sec}[1]{Sec.~\ref{sec:#1}}
\newcommand{\mscite}[1]{Ref.~\citen{#1}}
\newcommand{\mscites}[2]{Refs.~\citen{#1} and \citen{#2}}
\newcommand{\mmcite}[1]{Refs.~\citen{#1}}

\newcommand{\smfrac}[2]{\mbox{$\frac{#1}{#2}$}}
\newcommand{\ssum}[1]{{\textstyle\sum_{#1}\,}}

\newcommand{\conv}{\mathop{\mathrm{conv}}}
\newcommand{\rect}[2]{\mbox{$[\![$}#1,#2\mbox{$]\!]$}}

\newcommand{\tr}{\mathop{\mathrm{tr}}\nolimits}
\newcommand{\trdist}[2]{\mbox{$\smfrac{1}{2}\left\|\,#1-#2\,\right\|_1$}}

\newcommand{\verythinlines}{\linethickness{.15pt}}

\newcommand{\A}{{\rm A}}
\newcommand{\B}{{\rm B}}

\newcommand{\dAin} {d_{\A}^{\rm in}}
\newcommand{\dBin} {d_{\B}^{\rm in}}
\newcommand{\dAout}{d_{\A}^{\rm out}}
\newcommand{\dBout}{d_{\B}^{\rm out}}

\newcommand{\rsp}{{\mbox{\scriptsize\sc rsp}}}
\newcommand{\hsw}{{\mbox{\scriptsize\sc hsw}}}

\renewenvironment{proof}[1][]{%
\par\addvspace{12pt plus3pt minus3pt}\global\logotrue%
\noindent{\bf Proof#1.\hskip.5em}\ignorespaces}{%
        \par\iflogo\vskip-\lastskip
        \vskip-\baselineskip\prbox\par
        \addvspace{12pt plus3pt minus3pt}\fi}

\begin{document}

\markboth{Andrew M.\ Childs, Debbie W.\ Leung, and Hoi-Kwong Lo}
         {Two-way quantum communication channels}

\title{Two-way quantum communication channels}

\author{ANDREW M. CHILDS}

\address{Institute for Quantum Information \\
         California Institute of Technology \\
         Pasadena, CA 91125, USA \\
         amchilds@caltech.edu}

\author{DEBBIE W. LEUNG}

\address{Institute for Quantum Information \\
         California Institute of Technology \\
         Pasadena, CA 91125, USA \\
         wcleung@cs.caltech.edu}

\author{HOI-KWONG LO}

\address{Center for Quantum Information and Quantum Control, \\
         Department of Electrical \& Computer Engineering, \\
         and Department of Physics \\
         University of Toronto \\
         Toronto, Ontario M5S 3G4, Canada \\
         hklo@comm.utoronto.ca}

\maketitle


\begin{abstract}
We consider communication between two parties using a bipartite
quantum operation, which constitutes the most general quantum
mechanical model of two-party communication.  We primarily focus on
the simultaneous forward and backward communication of classical
messages.  For the case in which the two parties share unlimited prior
entanglement, we give inner and outer bounds on the achievable rate
region that generalize classical results due to Shannon.  In
particular, using a protocol of Bennett, Harrow, Leung, and Smolin, we
give a one-shot expression in terms of the Holevo information for the
entanglement-assisted one-way capacity of a two-way quantum channel.
As applications, we rederive two known additivity results for one-way
channel capacities: the entanglement-assisted capacity of a general
one-way channel, and the unassisted capacity of an
entanglement-breaking one-way channel.
\end{abstract}

\section{Introduction}

Suppose two parties, Alice and Bob, wish to exchange information.  To
do so, they must be connected by some physical interaction, or in
information-theoretic language, a channel.  One of the main problems
of information theory is to determine the maximum rates (i.e., the
capacities) for communication through such a channel.

A particularly well-studied type of interaction is the {\em one-way
channel}, which models transporting some carrier of information from a
fixed sender (say, Alice) to a fixed receiver (say, Bob), with the
state encoding the information possibly modified in some way during
transit.  In other words, a one-way channel is a formal change of
ownership of the state together with a state change.

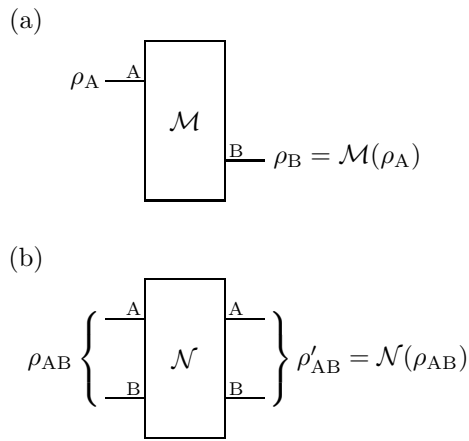
\begin{figure}[b]
\setlength{\unitlength}{1.5pt}
\begin{center}
\begin{picture}(117,110)
\put(0,40){\makebox(10,10){(b)}}
\put(5,5){\makebox(20,30){$\rho_{\A\B} \left\{\rule{0pt}{20pt}\right.$}}
\put(65,5){\makebox(52,30)
     {$\left.\rule{0pt}{20pt}\right\}\rho_{\A\B}'={\N}(\rho_{\A\B})$}}
\put(35,0){\framebox(20,40){$\N$}}
\put(25,10){\line(1,0){10}}
\put(55,10){\line(1,0){10}}
\put(25,30){\line(1,0){10}}
\put(55,30){\line(1,0){10}}
\put(30,30){\makebox(5,5){\scriptsize$\A\,$}}
\put(30,10){\makebox(5,5){\scriptsize$\B\,$}}
\put(55,30){\makebox(5,5){\scriptsize$\,\A$}}
\put(55,10){\makebox(5,5){\scriptsize$\,\B$}}
\put(0,100){\makebox(10,10){(a)}}
\put(15,85){\makebox(10,10){$\rho_\A$}}
\put(66,66){\makebox(40,10){$\rho_\B={\M}(\rho_\A)$}}
\put(25,90){\line(1,0){10}}
\put(35,60){\framebox(20,40){$\M$}}
\put(55,70){\line(1,0){10}}
\put(30,90){\makebox(5,5){\scriptsize$\A\,$}}
\put(55,70){\makebox(5,5){\scriptsize$\,\B$}}
\end{picture}
\end{center}
\caption{(a) One-way and (b) two-way quantum channels.}
\label{fig:channels}
\end{figure}

Classical one-way channels, introduced by Shannon \cite{Sha48}, are
stochastic maps on probability distributions.  More generally, in
quantum mechanics, evolution is described by a quantum operation,
i.e., a trace-preserving, completely positive map on quantum states
represented by density matrices (positive semidefinite matrices of
unit trace).  Figure \ref{fig:channels}(a) shows a one-way quantum
channel described by the quantum operation $\M$.  Alice prepares an
input state $\rho_\A$ which is transported to Bob as the output state
$\rho_\B = {\M}(\rho_\A)$.  The capacities to transmit classical
\cite{Holevo73,HJSWW96,Hol98,SW97,BSST99,BSST01} and quantum
\cite{Llo97,BNS98,Shor02a,Dev03} messages through such channels have
been extensively studied.

Although a one-way channel is an intuitive model for communication, it
is only a special case of the possible interactions between Alice and
Bob.  The most general interaction is a joint quantum operation $\N$
acting on the joint Hilbert space of the two parties.  If they are in
possession of a joint density matrix $\rho_{\A\B}$, then the action of
the channel produces a joint output state
$\rho_{\A\B}'={\N}(\rho_{\A\B})$.  In other words, Alice and Bob each
provide an input to the two-way channel, which evolves their inputs
jointly to produce a joint output shared by the two parties.  Such a
{\em two-way quantum channel} is shown in Fig.~\ref{fig:channels}(b).

The two-way quantum channel model is the most general setting for
two-party communication.  For example, one-way channels are simply
two-way channels with a zero-dimensional input for Bob and a
zero-dimensional output for Alice (or equivalently from an operational
standpoint, channels that discard Bob's input and give Alice a fixed
output).

Another subclass of two-way quantum channels consists of the classical
two-way channels.  Such channels were also first studied by Shannon
\cite{Sha61}, who gave outer and inner bounds on their capacity
regions.  Shannon's bounds were subsequently improved by many others
(see for example
\mmcite{Meu77,Due79,EC80,Sch82,Sch83,Han84,ZBS86,HW89}).

In the quantum setting, another natural special class of two-way
channels is the set of bipartite unitary interactions acting on
systems of fixed dimension.  The capacity question in this setting was
formalized and studied in \mscite{BHLS02}, and further capacity
expressions were subsequently found in \mscites{H03}{HL04}.  For such
channels, the problem of bidirectional communication is closely
related to the problem of generating entanglement
\cite{BHLS02,LHL03,BS03,H03,HL04}.
Generalizing to allow systems whose input and output dimensions are
different, one finds an especially simple class of interactions, the
{\em quantum feedback channels} \cite{igor}.  Such channels take no
input from Bob and evolve Alice's input into a state shared between
her and Bob.


A unifying viewpoint is that any two-way quantum channel can be viewed
as an isometry of the two input states to three output states,
discarding the third part of the system to an inaccessible
environment.  A variety of subclasses can be obtained simply by
changing the dimensions of the terminals.

The various subclasses of two-way channels can exhibit remarkably
different properties.  For example, any (nonlocal) two-way unitary
channel can communicate in either direction and generate entanglement
at a nonzero rate \cite{BHLS02}, while in each of the other examples,
some of these tasks are known to be impossible.  Because of the wide
variety of possible subclasses of two-way channels, and because
calculating capacities is known to be difficult for several of the
possibilities, we do not expect especially tight capacity results for
the general two-way channel.

A communication protocol using two-way channels may yield classical or
quantum communication in either or both directions.  It can also
create or consume classical or quantum correlations as auxiliary
resources.
In particular, providing Alice and Bob with enough free entanglement
considerably simplifies and unifies the study of communication
capacities of one-way quantum channels \cite{BSST99,BSST01} as well as
two-way channels \cite{BHLS02,BS03,H03,HL04}.

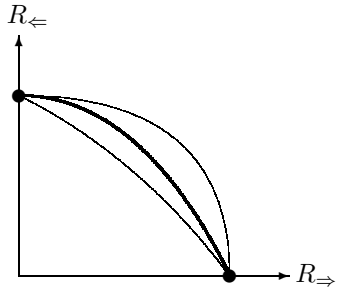
\begin{figure}[t]
\setlength{\unitlength}{0.8mm}
\begin{center}
\begin{picture}(60,55)(5,5)
\put(10,10){\vector(1,0){45}}
\put(10,10){\vector(0,1){40}}

\verythinlines
\qbezier(10,40)(45,40)(45,10)
\qbezier(10,40)(30,30)(45,10)

\thicklines
\qbezier(10,40)(30,40)(45,10)

\thinlines
\put(45,10){\circle*{2}}
\put(10,40){\circle*{2}}
\put(56,9){\makebox{$R_\cra$}}
\put(8,52){\makebox{$R_\cla$}}
\end{picture}
\end{center}
\vspace*{-3ex}
\caption{The achievable region for two-way classical communication at
fixed $R_{\rm e}$, with outer and inner bounds.}
\label{fig:capacityregion}
\end{figure}

In this paper, we will mostly be considering the set of achievable
rates of classical communication $R_\cra$ from Alice to Bob and
$R_\cla$ from Bob to Alice, and the rate of producing entanglement
$R_{\rm e}$ (which can either be positive, indicating that
entanglement is produced, or negative, indicating that entanglement is
consumed).
The set of achievable rates $(R_\cra,R_\cla,R_{\rm e})$ forms a
three-dimensional region.  The boundary of this region represents the
set of best achievable rates.  
Figure \ref{fig:capacityregion} shows a schematic diagram of the
achievable rates of classical communication $(R_\cra,R_\cla)$ at fixed
$R_{\rm e}$.  
Typically, there is a tradeoff between how much information Alice can
send to Bob and how much information Bob can send to Alice.
The end points of the optimal tradeoff curve, where $R_\cla = 0$ or
$R_\cra = 0$, are called the {\em one-way capacities} of the two-way
channel.
The cases in which $R_{\rm e}=0$ and $R_{\rm e} \ra -\infty$, called
{\em entanglement-unassisted} (or simply unassisted) and {\em
entanglement-assisted}, respectively, are of particular interest.

For the case of entanglement-assisted one-way classical communication
by unitary two-way channels, \mscite{BHLS02} gave a simplified
single-letter expression for the capacity, showed that it is additive,
and gave a protocol for achieving it.
These results were subsequently extended in \mscite{H03} to the case
in which a fixed amount of entanglement is consumed or generated (a
scenario dubbed {\em finite entanglement assistance}).  In addition,
tradeoff curves $\{(R_\cra,0,R_{\rm e})\}$ for one-way classical
communication were related to analogous curves $\{(R_\ra,0,R_{\rm
e})\}$ for quantum communication.
In \mscite{HL04}, the entire three-dimensional achievable regions for
finite entanglement-assisted bidirectional classical or quantum
communication were related.

In this paper, we consider bidirectional classical communication using
a two-way quantum channel in the general (not necessarily unitary)
case, possibly allowing entanglement assistance.  We begin in
\sec{prelim} by introducing some basic notation and formalizing the
notions of a protocol and the capacities it achieves.
In \sec{2wayeacc}, we extend techniques from \mscite{BHLS02} to
provide inner and outer bounds on the achievable region.  The outer
bound generalizes Shannon's bounds \cite{Sha61} to the quantum
setting.  The inner bound is an extension of the protocol of
\mscite{BHLS02} for simultaneous two-way communication, and also
reduces to Shannon's bound in the classical case.  Furthermore, the
bounds meet on the axes (as depicted in
Fig.~\ref{fig:capacityregion}), giving a formula for the one-way
capacity of a two-way quantum channel (extending the unitary result of
\mscite{BHLS02}).
In \sec{appl}, we describe two immediate applications of the inner and
outer bounds.  First, as a simple consequence of the fact that the
inner and outer bounds meet on the axes, we recover the additivity of
the entanglement-assisted classical capacity of one-way channels
\cite{BSST01}.  We also provide a simple operational derivation of the
additivity of the unassisted classical capacity of an
entanglement-breaking channel, a special case of a result that was
first proved in \mscite{Shor02b}.  
Finally, in \sec{concl}, we discuss some open questions and directions
for future investigation.

\section{Preliminaries}
\label{sec:prelim}

In this section, we describe our framework and define the notation
used throughout the paper.  

An {\em ebit} refers to a unit of shared quantum correlation, as
quantified by an EPR pair of qubits
$|\Phi\>_{\A\B}:=\frac{1}{\sqrt{2}}\sum_{x=0}^1 |x\>_{\A} |x\>_{\B}$,
with the density matrix $\Phi:=|\Phi\>\<\Phi|$.  Throughout the paper,
we omit the tensor product symbol, $\otimes$, if no confusion will
arise.
The functions $\exp$ and $\log$ are always base $2$.
The positive and negative parts of a real number $x$ are written as
$x^{\pm}$: for $x \ge 0$, $x^+ := x$ and $x^- := 0$; for $x < 0$, $x^+
:= 0$ and $x^- := -x$.
We adopt the convention that a quantum operation $\N$ acts on an
ensemble of quantum states $\E=\{p_i,\rho_i\}$ by acting on each state
individually (preserving its probability), i.e., $\N \E = \{p_i,
\N(\rho_i)\}$.

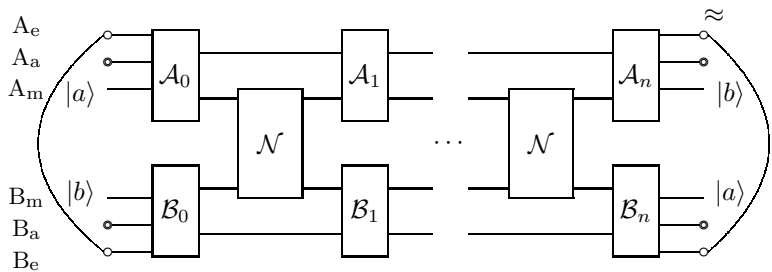
\begin{figure}[b]
\centering
\setlength{\unitlength}{0.6mm}
\begin{picture}(185,70)
{\small
\put(7,57){\makebox(10,8){$\A_{\rm e}$}}
\put(7,50){\makebox(10,8){$\A_{\rm a}$}}
\put(7,43){\makebox(10,8){$\A_{\rm m}$}}
\put(7,19){\makebox(10,8){$\B_{\rm m}$}}
\put(7,12){\makebox(10,8){$\B_{\rm a}$}}
\put(7,05){\makebox(10,8){$\B_{\rm e}$}}
}
\put(18,19){\makebox(12,10){$|b\>$}}
\put(30,11){\circle{2}}
\put(31,11){\line(1,0){9}}
\put(30,17){\circle{1}}
\put(30,17){\circle{2}}
\put(31,17){\line(1,0){9}}
\put(30,23){\line(1,0){10}}
\put(40,10){\framebox(10,20){$\cB_0$}}
\put(18,41){\makebox(12,10){$|a\>$}}
\put(30,47){\line(1,0){10}}
\put(31,53){\line(1,0){9}}
\put(30,53){\circle{1}}
\put(30,53){\circle{2}}
\put(31,59){\line(1,0){9}}
\put(30,59){\circle{2}}
\put(40,40){\framebox(10,20){$\cA_0$}}
\put(59,23){\framebox(14,24){$\N$}}
\put(50,15){\line(1,0){32}}
\put(50,25){\line(1,0){9}}
\put(73,25){\line(1,0){9}}
\put(82,10){\framebox(10,20){$\cB_1$}}
\put(50,45){\line(1,0){9}}
\put(73,45){\line(1,0){9}}
\put(50,55){\line(1,0){32}}
\put(82,40){\framebox(10,20){$\cA_1$}}
\put(92,15){\line(1,0){10}}
\put(92,25){\line(1,0){10}}
\put(92,45){\line(1,0){10}}
\put(92,55){\line(1,0){10}}
\put(101,30){\makebox(10,10){$\cdots$}}
\put(110,15){\line(1,0){32}}
\put(110,25){\line(1,0){9}}
\put(110,45){\line(1,0){9}}
\put(110,55){\line(1,0){32}}
\put(133,25){\line(1,0){9}}
\put(133,45){\line(1,0){9}}
\put(119,23){\framebox(14,24){$\N$}}
\put(142,10){\framebox(10,20){$\cB_n$}}
\put(142,40){\framebox(10,20){$\cA_n$}}
\put(162,19){\makebox(12,10){$|a\>$}} 
\put(162,17){\circle{1}}
\put(162,17){\circle{2}}
\put(152,17){\line(1,0){9}}
\put(162,11){\circle{2}}
\put(152,11){\line(1,0){9}}
\put(152,23){\line(1,0){10}}
\put(162,41){\makebox(12,10){$|b\>$}}
\put(152,47){\line(1,0){10}}
\put(152,53){\line(1,0){9}}
\put(162,53){\circle{2}}
\put(162,53){\circle{1}}
\put(152,59){\line(1,0){9}}
\put(162,59){\circle{2}}

\put(162,62){$\approx$}

\verythinlines
\qbezier(29,11)(0,35)(29,59)
\qbezier(163,11)(190,35)(163,59)
\end{picture}
\caption{Circuit representation of a general two-way channel
protocol.}
\label{fig:protocol}
\end{figure}

Our framework for communication using two-way channels extends the
model of \mscites{BHLS02}{HL04}.  
A general two-way quantum channel $\N$ has two inputs of dimensions
$\dAin, \dBin$ and two outputs of dimensions $\dAout, \dBout$.  
A general protocol $\cP_n$ for two-way classical communication with
$n$ uses of $\N$ consists of $n$ alternating steps of local processing
and application of $\N$ followed by a final step of local processing,
as depicted in Fig.~\ref{fig:protocol}.
Let $a,b$ be the respective messages to be communicated from Alice to
Bob and vice versa, and let $\cA_k,\cB_k$ be the local operations of
Alice and Bob between the $k$th and $(k+1)$th use of $\N$.
The operations $\cA_k,\cB_k$ may be arbitrary, but without loss of
generality, we can assume that all ancillas are present at the
beginning of the protocol and that all measurements are performed
coherently, with no systems discarded by $\cA_k,\cB_k$.  In other
words, all of $\cA_k,\cB_k$ can be assumed to be unitary.
Alice's initial operation $\cA_0$ has three input systems: $\A_{\rm
m}$, which contains the message $a$; $\A_{\rm a}$, which contains a
local ancilla; and $\A_{\rm e}$, which is maximally entangled with
$\B_{\rm e}$.
The operation $\cA_0$ converts $\A_{{\rm m},{\rm a},{\rm e}}$
unitarily into two systems: $\A$, which is the input to $\N$, and
$\A'$, which is not acted on by $\N$.
Each of $\cA_k$ for $k=1,{\ldots},n-1$ has two inputs $\A,\A'$ and two
outputs, which for simplicity will also be labeled as $\A,\A'$,
although in general they may have different dimensions than the input
systems.  The operation $\cA_k$ changes the dimensions of $\A,\A'$ if
$\dAin \neq \dAout$.
Finally, $\cA_t$ converts $\A,\A'$ to three output systems: $\A_{\rm
m}$, whose reduced state represents Bob's message for Alice; $\A_{\rm
a}$, which is an arbitrary ancillary system; and $\A_{\rm e}$, which
is nearly maximally entangled with $\B_{\rm e}$.
The situation is analogous on Bob's side. 

The goal is to find a family of protocols $\{\cP_n: n \in \bbN\}$,
where protocol $\cP_n$ employs $n$ uses of $\N$, that will perform the
communication task with accuracy that can be made arbitrarily good by
taking $n$ sufficiently large---the output state of $\A_{\rm
m},\B_{\rm m}$ should be close to $|b\>\<b| \ot |a\>\<a|$, the state
of $\A_{\rm e},\B_{\rm e}$ should be nearly maximally entangled, and
$\A_{\rm a},\B_{\rm a}$ should be nearly disentangled from $\A_{{\rm
m},{\rm e}},\B_{{\rm m},{\rm e}}$ (but can depend on the messages
$a,b$).  
The input dimensions of $\A_{\rm m},\B_{\rm m}$ determine the
communication rates, while the change of the dimensions of $\A_{\rm
e},\B_{\rm e}$ determines the amount of entanglement consumed or
generated by $\cP_n$.

To quantify the proximity of two quantum states $\rho,\sigma$, we use
the trace distance $\trdist{\rho}{\sigma}$, where $\|X\|_1 :=
\tr\sqrt{X^\dag X}$.
For example, for two pure states $|\alpha\>, |\beta\>$,
  $\trdist{|\alpha\>\<\alpha|}{|\beta\>\<\beta|} =
  \epsilon \, \Leftrightarrow \, |\<\alpha|\beta\>|^2 = 1-\e^2$.


We are now ready to define the achievable region: 
\begin{definition}[Achievable rates]
\label{def:ach}
We say that $(R_\cra,R_\cla,R_{\rm e})$ is achievable if there is a
sequence of protocols $\{ \cP_n \}$, together with asymptotically
vanishing sequences $\{\delta_n\},\{\epsilon_n\}$ of nonnegative
numbers and sequences of integers
$\{r_{\cra,n}\},\{r_{\cla,n}\},\{r_{{\rm e},n}\}$ such that
\ba
       r_{\cra,n} &\geq n(R_\cra-\delta_n) \,, \\
       r_{\cla,n} &\geq n(R_\cla-\delta_n) \,, \\
  r_{{\rm e},n}^+ &\geq n(R_{\rm e}^+-\delta_n) \,, \\
  r_{{\rm e},n}^- &\leq n(R_{\rm e}^-+\delta_n) \,,
\ea
and for all messages $a \, {\in} \, \{0,1\}^{r_{\cra,n}}$, $b \, {\in}
\, \{0,1\}^{r_{\cla,n}}$ the following success criteria hold.
Let
\be
  \rho_n^{ab} := \cP_n (|0\>\<0|_{\A_{\rm a}} |0\>\<0|_{\B_{\rm a}}
                    \ot |a\>\<a|_{\A_{\rm m}} |b\>\<b|_{\B_{\rm m}} 
                    \ot \Phi^{\ot r_{{\rm e},n}^-}_{\A_{\rm e}\B_{\rm e}})
\,,
\ee
and let 
$\rho^{ab}_{n,\A_{\rm m}\B_{\rm m}}$, 
$\rho^{ab}_{n,\A_{\rm e}\B_{\rm e}}$ be its reductions to the message
and entanglement subspaces, respectively. 
Then the success criteria are
\ba
  \trdist{\rho^{ab}_{n,\A_{\rm m}\B_{\rm m}}}
         {|b\>\<b|_{\A_{\rm m}} |a\>\<a|_{\B_{\rm m}}}
  &\leq \epsilon_n \,,
\label{eq:cc-condition} \\
  \trdist{\rho^{ab}_{n,\A_{\rm e}\B_{\rm e}}}
         {\Phi^{\ot r_{{\rm e},n}^+}_{\A_{\rm e}\B_{\rm e}}}
  &\leq \epsilon_n
\,.
\ea
\end{definition}

The achievable region is {\em convex} by convex combination (i.e.,
time sharing) of protocols.  
It is {\em monotone} since resources can always be discarded to give a
lower rate of communication, lower rate of entanglement production,
or higher rate of entanglement consumption.

Any quantum protocol for communicating classical information concludes
by decoding the quantum state to produce a classical output.  The
final decoding can be viewed as a measurement of the final state,
which in our model is implemented unitarily.  The general problem of
extracting classical information encoded in a quantum state (i.e.,
learning the index $i$ in a random draw from an ensemble
$\E=\{p_i,\rho_i\}$ of quantum states) has been thoroughly studied in
this context \cite{Holevo73,HJSWW96,Hol98,SW97}.
The accessible information of $\E$, denoted $I_{\rm acc}(\E)$, is
defined to be the maximum mutual information between the label $i$ and
the outcome of any possible measurement.  It is upper
bounded as \cite{Holevo73}
\be
  I_{\rm acc}(\E) \le \chi(\E)
\label{eq:holevobound}
\,,
\ee
where $\chi(\E)$ denotes the {\em Holevo information} of $\E$,
\be
  \chi(\E) := S\bigg(\sum_i p_i \rho_i\bigg) - \sum_i p_i S(\rho_i)
\label{eq:holevo}
\,.
\ee
Here $S(\rho):=-\tr{\rho \log \rho}$ is the von Neumann entropy of
$\rho$.

In fact, by appropriate encoding, the Holevo information turns out to
be asymptotically achievable in the following sense.  Consider a {\em
CQ channel}, i.e., a one-way channel with a classical input $i$ giving
rise to a corresponding quantum output state $\rho_i$.  Then we have

\begin{theorem}[HSW Theorem \cite{HJSWW96,Hol98,SW97}]
\label{thm:hsw}
The capacity of the CQ channel $\{i \ra \rho_i\}$ is given by
$ \sup_{\{p_i\}} \chi(\{p_i, \rho_i\})$.
\end{theorem}

More specifically, for any fixed input probability distribution
$\{p_i\}$, let $\E=\{p_i,\rho_i\}$ denote the corresponding ensemble.
In the limit of large $n$, there is a set of
$\exp(n\chi(\E)-\delta_n)$ states $\rho_{i_1} \ot \rho_{i_2} \ot
\cdots \ot \rho_{i_n}$ that can be distinguished with error bounded by
some $\e_n$, where $\delta_n, \e_n \to 0$ as $n \to \infty$.
This is proved by showing that, with nonzero probability, a random
code (in which each tensor component of each codeword is drawn
independently from $\E$) can be decoded with vanishing error.

The HSW theorem can be applied to an arbitrary one-way channel by
viewing the states $\rho_i$ as the possible outputs of the channel.
However, for a general two-way channel, the coding problem is
complicated somewhat by the fact that the input and output ensembles
are bipartite.  In this case, communication in one direction is
typically affected by the input to the channel from the opposite
direction.  Thus, we will need to prove a modified version of the HSW
theorem when we derive protocols for bidirectional communication.

\section{Entanglement-assisted capacity of two-way quantum channels}
\label{sec:2wayeacc}

In this section, we present inner and outer bounds on the classical
capacity region of an entanglement-assisted two-way quantum channel.  

Let $\cP_n$ be an arbitrary protocol that employs $n$ uses of the
two-way channel $\N$.
We assume without loss of generality that Alice and Bob retain a copy
of their input messages throughout the protocol.  Then, after $t$ uses
of $\N$ followed by $\cA_t \ot \cB_t$, Alice and Bob possess a joint
state drawn from an ensemble
\be
  \E^{(t)} = \{p_{ab}, |a\>\<a| \ot \rho^{(t)}_{ab} \ot |b\>\<b|\}
\label{eq:ensemble}
\ee
indexed by the classical messages $a,b$ to be communicated, where
$\E^{(t)}$ is completely determined by $\cP_n$.

For a bipartite ensemble $\E$ over systems $\A\A'$ and $\B\B'$, we
define the {\em local Holevo information} of $\B\B'$ and $\A\A'$,
respectively, as
\ba
  \chi_\cra(\E) &:= \chi(\tr_{\A \A'} \E) \,, \\
  \chi_\cla(\E) &:= \chi(\tr_{\B \B'} \E)
\,.
\ea
Following earlier work \cite{BHLS02}, our analysis will be based on
examining the local Holevo information $\chi_\cra(\E^{(t)})$,
$\chi_\cla(\E^{(t)})$ for $t=0,1,\ldots,n$ during the course of an
$n$-use protocol.  
By the joint entropy theorem (see for example Eq.~(1.58) in
\mscite{Nielsen00}), for an ensemble $\E^{(t)}$ of the form of
\eq{ensemble} (i.e., with local copies of the classical messages), the
local Holevo information can be rewritten as
\ba
  \chi_\cra(\E^{(t)}) &= H(\{p_b\}) 
                         + \sum_b p_b \, \chi(\E_b^{(t)}) \,,
\label{eq:chibob} \\
  \chi_\cla(\E^{(t)}) &= H(\{p_a\}) 
                         + \sum_a p_a \, \chi(\E_a^{(t)})
\label{eq:chialice}
\,, 
\ea
where $p_a := \sum_b p_{ab}$ and $p_b := \sum_a p_{ab}$ are the
marginal distributions and
  $\E_b^{(t)} := \{ p_{a|b}, \tr_{\A\A'} \rho_{ab}^{(t)} \},
   \E_a^{(t)} := \{ p_{b|a}, \tr_{\B\B'} \rho_{ab}^{(t)} \}$
are the ensembles for Bob and Alice conditioned on their known inputs
$b$ and $a$, respectively, where $p_{\cdot|\cdot}$ denotes conditional
probability.

Equations (\ref{eq:chibob}) and (\ref{eq:chialice}) give natural
interpretations of the local Holevo information: for example, for Bob,
it is the sum of the information about $b$ that he already knows and
the information about $a$ obtainable from $\E^{(t)}_b$, averaged over
$b$.
Removing the information that the sender already knows gives a
quantity that is useful for obtaining bounds.  Thus, we define the
{\em readjusted local Holevo information} as 
\ba
  \bar \chi_\cra(\E^{(t)}) &:= \sum_b p_b \, \chi(\E_b^{(t)}) 
\label{eq:barchibob} \,, \\
  \bar \chi_\cla(\E^{(t)}) &:= \sum_a p_a \, \chi(\E_a^{(t)})
\label{eq:barchialice}
\,. 
\ea

\subsection{Additive outer bound}
\label{sec:addoutbdd}

In this section, we obtain an additive outer bound on the capacity
region for entanglement-assisted bidirectional classical communication
using a two-way quantum channel.

Consider the differences in local Holevo information induced by an
application of $\N$ to an ensemble $\E$: 
\ba
  \Delta\chi_\cra(\E) &:= \chi_\cra(\N\E) - \chi_\cra(\E) 
                        = \bar \chi_\cra(\N\E) - \bar \chi_\cra(\E) 
\,,
\\
  \Delta\chi_\cla(\E) &:= \chi_\cla(\N\E) - \chi_\cla(\E)
                        = \bar \chi_\cla(\N\E) - \bar \chi_\cla(\E) 
\,.
\ea
Let $\rect{x}{y}$ denote the region $[0,x] \times[0,y] \subset
\bbR^2$, and let $\conv(\cdot)$ denote the convex hull.  
In terms of these quantities, we have the following outer bound on the
achievable region:

\begin{theorem}
\label{thm:outerbound}
  If $(R_\cra,R_\cla,-\infty)$ is achievable, then 
\be
  (R_\cra,R_\cla) \in 
    \conv \{ \rect{\Delta\chi_\cra(\E)}{\Delta\chi_\cla(\E)}:
    {\rm arbitrary~} \E\}
\,.
\ee
\end{theorem}

\begin{proof}
For any $\cP_n$, the Holevo bound on the accessible information,
\eq{holevobound}, implies that the number of bits that can be
faithfully transmitted forward and backward are no more than $\bar
\chi_\cra(\E^{(n)})$ and $\bar \chi_\cla(\E^{(n)})$ respectively.
In other words, 
\ba
\label{eq:holevobdd0}
  R_\cra &\leq \smfrac{1}{n} \bar \chi_\cra(\E^{(n)}) \,,  \\
  R_\cla &\leq \smfrac{1}{n} \bar \chi_\cla(\E^{(n)}) \,.
\label{eq:holevobdd1}
\ea
Expressing $\bar \chi_\cra(\E^{(n)})$ as a telescopic sum and 
using the fact $\bar \chi_\cra(\E^{(0)}) = 0$, 
\ba
  \bar \chi_\cra(\E^{(n)}) 
  &= \sum_{t=1}^n
       [\bar \chi_\cra (\E^{(t)}) - \bar \chi_\cra (\E^{(t-1)})] \,. 
\\
  &= \sum_{t=1}^n \Delta \chi_\cra (\E^{(t-1)})  \,,   
\label{eq:tel1}
\ea
where \eq{tel1} comes from the fact that $\E^{(t)} = (\cA_t \ot \cB_t)
\N \E^{(t-1)}$, and $\cA_t$, $\cB_t$ leave $\bar \chi_\cra$ invariant
so that $\bar \chi_\cra (\E^{(t)}) = \bar \chi_\cra (\N \E^{(t-1)})$.
Likewise,
\be
  \bar \chi_\cla(\E^{(n)}) = \sum_{t=1}^n \Delta \chi_\cla (\E^{(t-1)}) 
\,.
\label{eq:tel2}
\ee
Putting \eqssss{holevobdd0}{holevobdd1}{tel1}{tel2} together, we find
\ba 
	(R_\cra, R_\cla) &\in
	\rect{\smfrac{1}{n} \chi_\cra(\E^{(n)})}
             {\smfrac{1}{n} \chi_\cla(\E^{(n)})} \\
	&\subset
	\conv \{
	\rect{\Delta\chi_\cra(\E)}{\Delta\chi_\cla(\E)}:
        {\rm arbitrary} ~\E\} \,,
\ea 
which completes the proof.
\end{proof}

As a side remark, we can also keep track of the entanglement $E$ of
the ensemble $\E^{(t)}$ along with the local Holevo information.  We
define $\Delta E(\E):=E(\N\E) - E(\E)$, where $E(\{ p_i, \rho_i\}) :=
\sum_i p_i E(\rho_i)$ and $E$ is defined to be the distillable
entanglement if $R_{\rm e} \geq 0$ and the entanglement cost if
$R_{\rm e} \leq 0$.
Thus, we have the following: 
\begin{theorem}
  If $(R_\cra,R_\cla,R_{\rm e})$ is achievable, then 
\be
  (R_\cra,R_\cla,R_{\rm e}) \in 
    \conv \{ 
      \rect{\Delta\chi_\cra(\E)}{\Delta\chi_\cla(\E)} \times
      [-\Delta E(\E)^{-},\Delta E(\E)^{+}]: \text{arbitrary~} \E
	  \}
\,. 
\ee
\end{theorem}
\noindent
However, in the remainder of \sec{2wayeacc}, we will consider only the
case of unlimited entanglement assistance.

\subsection{Two-way communication protocols based on remote state
preparation}

Having derived outer bounds on the achievable region, we now turn to
inner bounds based on protocols that achieve particular rates of
communication.
The main idea is to generalize the protocol from Sec.\ 4.3 of
\mscite{BHLS02} (and see also generalizations in \mscites{H03}{H05}),
originally designed for one-way communication with two-way unitary
channels, to the general problem of two-way communication with
possibly nonunitary two-way channels.
The main ingredients of the protocol of \mscite{BHLS02} are the HSW
Theorem and {\em remote state preparation} (RSP), which allows Alice
to share states of her choice with Bob.
Our two-way protocol is based on two-way analogs of the HSW Theorem
and RSP.
In this section, we describe these tools, present the generalized
two-way protocol, and analyze its error rate and inefficiency.  In the
following section, we show that combining the general technique with a
particular method of remote state preparation gives an explicit inner
bound.

First, we describe a bidirectional version of the HSW Theorem.  Such a
tool is necessary since the effective channel through which Alice can
send signals to Bob depends on what input Bob is using to send signals
to Alice, and vice versa.  Consider any {\em two-way CQ channel},
i.e., a channel with two classical inputs, $i$ for Alice and $j$ for
Bob, giving rise to a joint quantum output state $\rho_{ij}$.  In this
setting, we have the following:

\begin{lemma}[Bidirectional HSW inner bound]
\label{lem:hsw-bidir}
For the two-way CQ channel $\{i,j \to \rho_{i,j}\}$, rates
$(R_\cra,R_\cla)$ satisfying
\be
  (R_\cra,R_\cla) \in
    \conv\{\rect{\bar\chi_\cra(\E)}{\bar\chi_\cla(\E)}: 
    \E=\{p_i q_j,\rho_{ij}\}\}
\ee
are achievable.
\end{lemma}

\begin{proof}[~(sketch)]
We omit the straightforward (but lengthy) generalization of the proof of
Theorem~\ref{thm:hsw} in \mscites{SW97}{Hol98}.  
The basic idea is as follows.
For each use of the two-way channel, Alice is unaware of Bob's input
(which defines the effective channel from Alice to Bob), but such
information {\em is} available for Bob in his decoding operation.  The
same holds for communication from Bob to Alice.
Then, it is possible to show that good random codes (chosen
independently by Alice and Bob) exist, and allow communication at the
above rates according to a packing lemma analogous to that in the
original proof.  
\end{proof}

Note that in this lemma, we have assumed that Alice and Bob choose
their signals independently, with encoding distributions $\{p_i\}$ for
Alice and $\{q_j\}$ for Bob, so that we can view Bob's encoding
distribution as inducing a distribution over CQ channels from Alice to
Bob, and vice versa.  However, this is not the most general encoding
distribution possible with many uses of the channel, so it is not
clear whether this inner bound for CQ channels can be exceeded.

To prepare ensembles for bidirectional communication, we consider
bipartite remote state preparation.  Here the goal is to prepare a
large number $n$ of states drawn from the bipartite ensemble
$\E=\{p_{ij},\rho_{ij}\}$ with each party knowing one of the labels
$i,j$.  Note that Alice's label may control Bob's portion of the
state as well as her own (indeed, the state may be entangled between
their respective systems), and similarly for Bob's label.

We will assume the existence of a (not necessarily optimal) protocol
for bipartite remote state preparation with known asymptotic classical
communication and entanglement costs $C_\cra,C_\cla$ and $C_{\rm e}$.
More specifically, suppose $n (C_\cra+\delta_n^\rsp)$ forward
classical bits, $n (C_\cla+\delta_n^\rsp)$ backward classical bits,
and $n (C_{\rm e}+\delta_n^\rsp)$ ebits are sufficient for Alice and
Bob to prepare a state drawn from $\E^{\ot n}$ with fidelity
$1-\e_n^\rsp$, such that $\delta_n^\rsp, \e_n^\rsp \ra 0$ as $n \to
\infty$.
The problem of optimizing such costs in general is quite difficult,
and has only been solved for very special cases of $\E$.  But assuming
the existence of such a protocol to prepare any particular ensemble of
the form $\E=\{p_i q_j, \rho_{ij}\}$ (where the labels are chosen
independently but the corresponding state may be arbitrary) at given
costs, a corresponding point can be attained in the achievable region.

\begin{lemma} 
\label{lem:looking-glass}
$(R_\cra,R_\cla,-\infty)$ is achievable for all
\ba
  (R_\cra,R_\cla) \in
    \conv \{ \rect{\Gamma_{\!\cra}(\E)}{\Gamma_{\!\cla}(\E)}: 
    &~ \E=\{p_a q_b,\rho_{ab}\} \text{~such~that~} \nn
    &~ C_\cra(\E) \leq \bar\chi_\cra(\N\E)-\Gamma_{\!\cra}(\E), \nn
    &~ C_\cla(\E) \leq \bar\chi_\cla(\N\E)-\Gamma_{\!\cla}(\E) \}
\,.
\ea
\end{lemma}

Here $\bar\chi_\cra(\N\E), \bar\chi_\cla(\N\E)$ represent achievable
forward and backward communication rates for the ensemble $\N\E$
(according to Lemma~\ref{lem:hsw-bidir}).
The quantities $\Gamma_{\!\cra}(\E), \Gamma_{\!\cla}(\E)$ thus
represent the amount of communication gained by one use of $\N$ on the
ensemble $\E$; that is, the communication rates of $\N\E$ minus the
communication costs of preparing $\E$. 

\begin{proof}
By convexity and monotonicity, we only need to show that 
  $R_\cra = \bar \chi_\cra(\N\E)-C_\cra(\E)$ and 
  $R_\cla = \bar \chi_\cla(\N\E)-C_\cla(\E)$
are achievable rates (given a sufficiently large amount of
entanglement assistance).  
We do this by giving a communication protocol achieving those rates
assuming the existence of an RSP protocol with the stated
communication costs.
Since the protocol is a generalization of that in Sec.\ 4.3 of
\mscite{BHLS02} for one-way communication, some readers may wish to
refer to the detailed description and schematic diagram therein.

The protocol is as follows. 
Alice and Bob preagree on sufficiently large values of $n$ and $k$ (to
be determined later) and proceed with the following protocol, using
$\N$ approximately $nk$ times to communicate $k$ messages $a_1, a_2,
\ldots, a_k$, each consisting of
  $n (R_\cra -\delta_n^\hsw-\delta_n^\rsp)$
bits, in the forward direction,
and $k$ messages $b_1, b_2, \ldots, b_k$, each consisting of
  $n (R_\cla -\delta_n^\hsw-\delta_n^\rsp)$
bits, in the backward direction.

\begin{enumerate}
\renewcommand{\labelenumi}{\arabic{enumi}.}
\item Using bipartite RSP, prepare a state $\rho_1$ from $\E^{\ot n}$
with fidelity at least $1-\e_n^\rsp$.  This requires $O(n)$ uses of
$\N$ to communicate $n (C_\cra(\E)+\delta_n^\rsp)$ bits from Alice to
Bob and $n (C_\cla(\E)+\delta_n^\rsp)$ bits from Bob to Alice.  (Note
that this initial communication is always possible if the appropriate
$\D\chi$ is positive for some ensemble (and otherwise, it is not
necessary).  Suppose that although $\D\chi_\cra>0$ for
some ensemble, $\D\chi_\cra \leq 0$ for all ensembles that can be created
at zero cost.  In this case the operation is {\em semicausal} from
Alice to Bob, and hence is also {\em semilocalizable}
\cite{BGNP01,ESW01}, meaning that it can be simulated by a local
operation by Bob, sending a quantum state to Alice, and a final local
operation by Alice.  But such an operation clearly has $\D\chi_\cra
\leq 0$ for all ensembles, which is a contradiction.  A similar
argument applies to $\D\chi_\cla$.)

\item Apply $\N^{\ot n}$ to $\rho_1$, which has been chosen such that
local measurements by Alice and Bob on $\N^{\ot n}(\rho_1)$ provide an
$n (\bar \chi_\cra(\N \E) - \delta_n^\hsw)$-bit message for Bob and an
$n (\bar \chi_\cla(\N \E) - \delta_n^\hsw)$-bit message for Alice with
probability at least $1 - \e_n^\rsp
- \e_n^\hsw$, according to Lemma~\ref{lem:hsw-bidir}.
Bob receives the message $a_1$ as well as the information needed to
perform RSP of $\rho_2 \in \E^{\ot n}$ in the next step.  Similarly,
Alice receives $b_1$ and together with the information she needs for
RSP in the next round.

\item Perform the $2\text{nd}, \ldots, k\text{th}$ rounds of RSP.
\end{enumerate}

Just as in \mscite{BHLS02}, Alice and Bob must know all of
$a_1,a_2,\ldots,a_k$ and $b_1,b_2,\ldots,b_k$ at the beginning of the
protocol, perform their measurements for the $k$th round of RSP to
obtain the RSP instructions for the $k$th round, encode them as part
of the message of the $(k-1)$th round of RSP, and proceed with their
part of the $(k-1)$th round of RSP, and so on, until the first round
RSP messages are generated and sent by the initial $O(n)$ uses of
$\N$.  For simplicity, we only consider non-interactive RSP protocols
(with only one round of communication from Alice to Bob and vice
versa), which will be sufficient for our applications.  (With one more
level of block coding, one should be able to use interactive RSP
protocols, but this would require a more detailed error analysis.)

Finally, we analyze the errors and inefficiencies to show
achievability of the rates.  Fix any desired $\delta,\e >0$.  The
above protocol employs $n(c+k)$ uses of $\N$ (for some constant $c$)
and communicates
  $nk (\bar \chi_\cra(\E) - C_\cra(\E) -\delta_n^\rsp-\delta_n^\hsw)$ 
and
  $nk (\bar \chi_\cla(\E) - C_\cla(\E) -\delta_n^\rsp-\delta_n^\hsw)$ 
bits forward and backward, respectively, with error 
  $k (\e_n^\hsw + \e_n^\rsp)$.
We can choose $k,n$ independently large enough so that
  $\smfrac{c}{k} + \delta_n^\rsp+\delta_n^\hsw < \delta$
and then increase $n$ if needed, to ensure 
  $k (\e_n^\hsw+\e_n^\rsp) \leq \e$.
\end{proof}

\subsection{Inner bound}

Explicit inner bounds for the achievable region can be obtained from
Lemma \ref{lem:looking-glass} together with known RSP protocols.  In
particular, we will make extensive use of the protocol for one-way RSP
of entangled states:

\begin{theorem}[RSP of entangled states \cite{BHLSW03}]
\label{thm:rsp}
  Asymptotically, the ensemble $\E = \{p_a, |\psi_a\>_{\A\B}\}$ can be
  prepared with a communication cost from Alice to Bob of
  $C_\cra(\E) = \chi_\cra(\E)$ and a rate of entanglement consumption
  of $C_{\rm e}(\E) = \sum_a p_a S(\tr_\A |\psi_a\>\<\psi_a|)$.  
\end{theorem}

Using this RSP protocol in the general inner bound of
Lemma~\ref{lem:looking-glass}, we find

\begin{theorem}
\label{thm:innerbound}
$(R_\cra,R_\cla,-\infty)$ is achievable if 
\be
  (R_\cra,R_\cla) \in 
    \conv\{\rect{\D\chi_\cra(\E)}{\D\chi_\cla(\E)}: 
    \E = \{p_a, \rho_a \} \otimes \{q_b, \eta_b\} \}
\label{eq:prodens}
\ee
In the ensemble of \eq{prodens}, the tensor product decomposition of
$\A\A'\B\B'$ may be arbitrary: Alice can prepare joint states of any
subspace of her system and Bob's, as can Bob, so long as the two
subspaces are disjoint.  In other words, we can have $\rho_a \in {\rm
X}, \eta_b \in {\rm Y}$ for any fixed decomposition $\A\A'\B\B' =
\rm{X}\ot\rm{Y}$ into arbitrary complementary subspaces.
\end{theorem}

\begin{proof}
Alice and Bob are each given the knowledge of $a$ and $b$.
By Theorem \ref{thm:rsp}, $C_\cra(\E) = \chi(\{p_a, \rho_a \})$ and
$C_\cla(\E) = \chi(\{q_b, \eta_b\})$.  Thus, the result follows from
Lemma \ref{lem:looking-glass}.
\end{proof}

When one of the ensembles is trivial, the protocol performs one-way
communication, with $\bar \chi_\cla(\E)=0$ or $\bar \chi_\cra(\E)=0$
as appropriate.
For example, in the former case, the rate of forward communication
  $R_\cra = \D\chi_\cra(\E)$
is achievable for arbitrary $\E$, so that the inner and outer bounds
meet.  Therefore, we find an expression for the entanglement-assisted
one-way forward capacity of a two-way quantum channel.
Similarly, we find $R_\cla = \Delta\chi_\cla(\E)$ for the one-way
backward capacity.
Indeed, this result is immediate from the fact that the protocol of
\mscite{BHLS02} for one-way communication applies unchanged even when
$\cal N$ is not unitary.  Thus we have the following:

\begin{corollary}[One-way capacity of a two-way channel]
\label{cor:lglass}
\ba
  R_\cra^{\max} &:= \sup\{R_\cra: (R_\cra,0,-\infty) \text{~achievable}\}
                = \sup_\E \Delta\chi_\cra(\E) \\
  R_\cla^{\max} &:= \sup\{R_\cla: (0,R_\cla,-\infty) \text{~achievable}\}
                = \sup_\E \Delta\chi_\cla(\E)
\ea
where the supremum is over all ensembles $\{p_i,\rho_{i,\A\A'\B\B'}\}$
with ancillary systems $\A',\B'$.
\end{corollary}

In particular, we have
\begin{corollary}
$R_{\cra,\cla}^{\max}$ are strongly additive.  In other words, for any
pair of two-way quantum channels $\N,\N'$, we have
  $R_\cra^{\max}(\N \ot \N') = R_\cra^{\max}(\N)+R_\cra^{\max}(\N')$ and 
  $R_\cla^{\max}(\N \ot \N') = R_\cla^{\max}(\N)+R_\cla^{\max}(\N')$.
\label{cor:additive}
\end{corollary}
\noindent

\subsection{Relation to Shannon's classical bounds}
\label{sec:shannon}

Both the inner and outer bounds given above reduce to Shannon's
bounds\cite{Sha61} in the case of a two-way classical channel (in
which case entanglement assistance clearly does not help).

Consider sending information through a two-way classical channel.
Suppose the input symbols $a,b$ appear with the joint probability
distribution $p_{ab}$.  Then the output symbols $a',b'$ appear with
the joint probability distribution $p_{ab} \, p_{a'b'|ab}$, where the
conditional probabilities $p_{a'b'|ab}$ define the channel.  Let
$I({\rm X};{\rm Y}|{\rm Z})$ denote the {\em conditional mutual
information},
\be
  I({\rm X};{\rm Y}|{\rm Z}) := H({\rm X}|{\rm Z}) - H({\rm X}|{\rm YZ})
\,, 
\ee
where $H({\rm X}|{\rm Y})$ denotes the conditional Shannon entropy of
${\rm X}$ given ${\rm Y}$.  In terms of the conditional mutual
information, Shannon proved the following inner and outer bounds on
the capacity of a two-way classical channel:

\begin{theorem}[Shannon \cite{Sha61}]
If $(R_\cra,R_\cla)$ is achievable, then
\be
  (R_\cra,R_\cla) \in
    \conv\{ \rect{I(\A;\B'|\B)}{I(\B;\A'|\A)}: 
            p_{ab} \textrm{~arbitrary} \}
\,.
\ee
Conversely, if
\be
  (R_\cra,R_\cla) \in
    \conv\{ \rect{I(\A;\B'|\B)}{I(\B;\A'|\A)}: p_{ab}=p_a q_b \}
\,,
\ee
then $(R_\cra,R_\cla)$ is achievable.
\end{theorem}

Now consider the corresponding outer and inner bounds from
Theorems~\ref{thm:outerbound} and \ref{thm:innerbound}.  Let $\E = \{
p_{ab}, |aa\>_{\A\A'}|bb\>_{\B\B'} \}$ where $|a\>,|b\>$ are mutually
orthogonal states on systems $\A,\B$ and as before, the senders retain
copies of their inputs in systems $\A',\B'$.  The action of the
two-way classical channel $\N$ on this ensemble is
\be
  \N(|a\>\<a| \ot |b\>\<b|) 
  = \sum_{a',b'} p_{a'b'|ab} \, |a'\>\<a'| \ot |b'\>\<b'|
\,. 
\ee
It is straightforward to compute
\ba
  \chi(\tr_{\A\A'} \E)   &= H(\B) \\
  \chi(\tr_{\A\A'} \N\E) &= H(\B\B') - H(\B\B'|\A\B)
                     = H(\B) + I(\B';\A|B) 
\,.
\ea
Therefore,
\ba
  \Delta\chi_\cra(\E) &= I(\A;\B'|\B)
\,.  
\ea
Thus, we see that in the classical case, the outer bound of
Theorem~\ref{thm:outerbound} and the inner bound of
Theorem~\ref{thm:innerbound} are identical to Shannon's outer and
inner bounds, respectively.

The equivalence to Shannon's bounds shows that in general, the bounds
of Theorems \ref{thm:outerbound} and \ref{thm:innerbound} are not
tight; even in the classical case, the inner bound may be exceeded
\cite{Due79,Sch82,Sch83,Han84} and the outer bound may not be
achievable \cite{ZBS86,HW89}.  Such results for classical channels
might provide insight into how the general (quantum) bounds could be
tightened.

\section{Additivity results for one-way channels}
\label{sec:appl}

Since one-way quantum channels are simply special cases of two-way
channels, it is possible to obtain results about one-way channels by
thinking of them as two-way channels.  In this section, we use such an
approach to rederive two previously known additivity results for
one-way channels: the entanglement-assisted capacity of an arbitrary
one-way quantum channel, and the entanglement-unassisted capacity of
an entanglement-breaking one-way quantum channel.

In this section, $\M$ denotes a one-way channel, and the classical
capacity is simply the maximum value of $R_\cra$ in the achievable
region (at fixed $R_{\rm e}$).  We use $R(\M)$ and $R^E(\M)$ to denote
the classical capacity with no entanglement assistance and unlimited
entanglement assistance, respectively.

\subsection{Entanglement-assisted capacity of one-way channels}

A general expression for $R^E(\M)$ in terms of the quantum mutual
information was found in \mscite{BSST01}.  Furthermore,
\mscite{BSST01} proved that $R^E$ is strongly additive, i.e.,
$R^E(\M_1 \ot \M_2) = R^E(\M_1) + R^E(\M_2)$ for any pair of (one-way)
quantum channels $\M_1,\M_2$.

The original proof of additivity used entropy inequalities to show
that the explicit expression for $R^E$ is indeed additive
\cite{BSST01,CA97}.  But specializing Corollary~\ref{cor:additive} to
one-way quantum channels provides an immediate alternative proof of
strong additivity.

These two proofs appear to be inequivalent.  The simplicity of proving
additivity via Corollary~\ref{cor:additive} seems to follow from the
structure of the protocol of \mscite{BHLS02} (or equivalently, that in
Lemma \ref{lem:looking-glass}).  The main idea of this protocol, to
borrow a resource and later regenerate some or more of it, has
recently found a number of applications in quantum information theory
\cite{family,H03,HL04}.  Such a protocol gives rise to a coding
structure very different from more standard, direct techniques, such
as those used in \mscite{BSST01}.  

{From} Corollary~\ref{cor:lglass}, we see that the capacity expression
of \mscite{BSST01} in terms of the quantum mutual information can be
written as a supremum of $\Delta\chi$.  It is not obvious simply by
looking at these two expressions that they are in fact equal.

\subsection{Unassisted capacity of one-way entanglement-breaking
channels}

We now turn our attention to unassisted classical communication using
a one-way channel.  In particular, we consider {\em
entanglement-breaking channels}, which are guaranteed to output a
state that is unentangled between the sender and the receiver.  Using
the framework of two-way channels, we will prove a special case of the
following result:
\begin{theorem}[Shor \cite{Shor02b}]
\label{thm:entbreak}
If $\M$ is an arbitrary one-way quantum channel and $\M'$ is an
entanglement-breaking one-way quantum channel, then 
\be
  R(\M \otimes \M') = R(\M) + R(\M') \,.
\ee
\end{theorem}

We will prove this result in the special case in which both $\M$ and
$\M'$ are entanglement-breaking.  In particular, this includes the
case $\M=\M'$, demonstrating the additivity of the Holevo capacity of
an entanglement-breaking channel.
The proof in terms of two-way channels for this special case is
significantly simpler.

As in Shor's proof \cite{Shor02b}, we use strong subadditivity of the
von Neumann entropy \cite{LR73} in various guises.  In particular, we
will use the following lemma:

\begin{lemma}
\label{lemma:star}
Let $\{\sigma_i\},\{\eta_i\}$ be sets of quantum states, and let
$\{p_i\}$ be a probability distribution.  Then
\be
  S(\ssum{i} p_i \, \s_i \ot \eta_i) 
  \ge
  S(\ssum{i} p_i \, \s_i) 
  + \ssum{i} p_i \, S(\eta_i)
\,.
\label{eq:plus}
\ee
\end{lemma}

We give two proofs of this lemma: an operational proof and a proof
that uses strong subadditivity directly.

\begin{proof}[~1 (Operational)]
Let $\E_1 = \{p_i, \s_i\}$ and $\E_2 = \{p_i, \s_i \ot \eta_i\}$.  
Since $\E_1$ can be obtained from $\E_2$ by discarding 
the second system, 
\ba
	0 & \leq \chi(\E_2) -\chi(\E_1) \\
	&= S(\ssum{i} p_i \, \s_i \ot \eta_i) - \ssum{i} p_i \, S(\eta_i) 
	 - S(\ssum{i} p_i \, \s_i )
\,, 
\ea
where the last line is obtained by using the definition \eq{holevo}
and the fact that $S(\s \ot \eta) = S(\s) + S(\eta)$.  
\end{proof}

\begin{proof}[~2 (Direct use of strong subadditivity)]
For the state 
$\r_{ABC} := \sum_i \, p_i \, \s_{i,A} \ot \eta_{i,B} \ot |i\>\<i|_C$, 
\ba 
  S_{ABC} &= H(\{p_i\}) + \ssum{i} p_i \, S(\s_i \ot \eta_i) \\
  S_{AB}  &= S(\ssum{i} p_i \, \s_i \ot \eta_i) \\
  S_{AC}  &= H(\{p_i\}) + \ssum{i} p_i \, S(\s_i) \\
  S_{A}   &= S(\ssum{i} p_i \, \s_i)
\,.
\ea
Equation (\ref{eq:plus}) then follows from the strong subadditivity
inequality $S_{ABC} + S_A \le S_{AB} + S_{AC}$.
\end{proof}

\begin{proof}[~of Theorem \ref{thm:entbreak} for \boldmath$\M$
entanglement-breaking]
The idea of the proof is to show that the states that can be output by
either channel are of no use in enhancing the capacity of the other
channel, and hence that the capacity of the joint channel is simply
the sum of the individual capacities.

For any entanglement-unassisted protocol that uses only
entanglement-breaking channels, we can rerun our proof of the outer
bound in \sec{addoutbdd} restricting to ensembles of separable states.
Thus we have an upper bound analogous to \eq{tel2},
\be
  R(\M) \leq 
  \max \{\Delta \chi_\cra (\E): \text{separable~} \E \} 
\label{eq:tel3}
\,.
\ee

Applying the two-way channel formalism to a one-way channel $\M$ with
input system $\A$ and output system $\B$, the most general input and
output ensembles are $\E_{\rm in} = \{p_i, \r_{i, \A' \A \B'}\}$ and
$\E_{\rm out} = \{p_i, \M(\r_{i, \A' \A \B'})\}$.
Without loss of generality, we can omit the system $\A'$.  This system
does not appear in the bound on the communication rate in terms of
$\Delta \chi_\cra$, and there is no entanglement to be stored in
$\A'$. 
Thus, the optimal input ensemble can be restricted to have the form
$\E_{\rm in} = \{p_i, \r_{i, \A\B'}\}$ with $\E_{\rm out} = \{p_i,
\M(\r_{i, \A \B'}) \}$.
 
We will be interested in three cases where the form of the input
ensemble is restricted to different extents.  
Let $\D \chi^S$, $\D \chi^P$, $\D \chi^0$ denote
$\sup_{\E_{\rm in}} [\chi(\E_{\rm out}) - \chi(\tr_{\A} \E_{\rm in})]$
for $\E_{\rm in}$ ranging over 
\ba
  \E_{\rm in}^S &:= \{p_i, \ssum{j} q_{ij} \, \r_{ij,\A} \ot \s_{ij,\B'}\} \,,
\label{eq:Sens}
\\
  \E_{\rm in}^P &:= \{p_i, \r_{i,\A} \ot \s_{i,\B'}\} \,,
\label{eq:Pens}
\\
  \E_{\rm in}^0 &:= \{p_i, \r_{i,\A}\}
\label{eq:0ens}
\ea
for separable ensembles, product ensembles, and ensembles with
$\chi=0$, respectively.
It is clear that $\D\chi^0 \le \D\chi^P \le \D\chi^S$; we will show
that $\D\chi^S \leq \D\chi^P \leq\D \chi^0$, so that in fact, all
three quantities are equal.

First we show that $\D\chi^S \leq \D\chi^P$.  For any separable
ensemble in the form of \eq{Sens}, let $\eta_{ij,\A} :=
\M(\r_{ij,\A})$.  Then we have
\ba
  \chi(\E_{\rm out}) - \chi(\tr_{\A} \E_{\rm in})
    &= S(\ssum{ij} \, p_i \, q_{ij} \, \s_{ij,\B'} \ot \eta_{ij,\A})
       -\ssum{i} p_i \, S(\ssum{j} q_{ij} \, \s_{ij,\B'}\ot \eta_{ij,\A}) \nn
    &\quad -S(\ssum{ij} p_i \, q_{ij} \, \s_{ij,\B'})
       +\ssum{i} p_i \, S(\ssum{j} q_{ij} \, \s_{ij,\B'}) \\
    &\le S(\ssum{ij} p_i \, q_{ij} \, \s_{ij,\B'} \ot \eta_{ij,\A})
       -\ssum{ij} p_i \, q_{ij} \, S(\s_{ij,\B'}\ot \eta_{ij,\A}) \nn
    &\quad -S(\ssum{ij} p_i \, q_{ij} \, \s_{ij,\B'})
       +\ssum{ij} p_i \, q_{ij} \, S(\s_{ij,\B'}) \leq  \D \chi^P
\ea
where Lemma \ref{lemma:star} has been applied to each term in $\sum_i
p_i \, S(\cdot)$.  Thus, a separable ensemble is no better than a
product ensemble. 

Now we show that $\D\chi^P \le \D\chi^0$.  For any product ensemble in
the form of \eq{Pens}, let $\eta_{i,\A} := \M(\r_{i,\A})$.  Then
\ba
  \chi(\E_{\rm out}) - \chi(\tr_{\A} \E_{\rm in})
    &= S(\ssum{i} p_i \, \s_{i,\B'} \ot \eta_{i,\A}) 
       -\ssum{i} p_i \, S(\s_{i,\B'}\ot \eta_{i,\A}) \nn
    &\quad -S(\ssum{i} p_i \, \s_{i,\B'})
      +\ssum{i} p_i \, S(\s_{i,\B'}) \\
  &\leq S(\ssum{i} p_i \, \eta_{i,\A}) 
     -\ssum{i} p_i \, S(\eta_{i,\A}) \leq \D \chi^0
\ea
where the inequality is due to subadditivity of $S(\sum_i p_i \,
\s_i \ot \eta_i)$ and additivity of $S(\s \ot \eta)$.  Thus a product
ensemble with $\chi \neq 0$ is no better than one with $\chi = 0$.

This argument shows that we can assume without loss of generality that
the input ensemble is of the form of \eq{0ens}.  But such an ensemble
costs nothing to create, so by using the protocol of Lemma
\ref{lem:looking-glass}, we see that the capacity $\D\chi^0$ can be
achieved for any ensemble $\E_{\rm in}^0$, and
\be
  R(\M) = \sup_{\E_{\rm in}^0} \D\chi^0
\,.
\ee

Finally, consider the capacity of the combined channel $\M \ot \M'$
where both $\M$ and $\M'$ are entanglement-breaking.  Without loss of
generality, we can assume that the channels act sequentially.  Each
channel can only produce separable output states, which by the above
argument are no better than states with $\chi=0$, which can be
produced at zero cost.  Therefore the capacity of the combined channel
is simply the sum of the individual capacities.
\end{proof}

\section{Open questions}
\label{sec:concl}

In this paper, we have established simple inner and outer bounds on
the entanglement-assisted classical capacity region of a two-way
quantum channel, and we have applied the framework of two-way channels
to rederive two previous additivity results for one-way channels.
However, since the two-way channel framework includes a wide variety
of disparate communication scenarios as special cases, this work
raises many more questions than it answers.

Calculating the capacity region for any particular channel can be a
challenging problem, and has not been done except in a few particular
special cases.  In fact, even computing the inner and outer bounds
given in this paper can be difficult, since the ancillary state spaces
$\A',\B'$ may be arbitrarily large, and we do not know that
low-dimensional ancillas are sufficient to achieve the capacity (even
in the unitary case \cite{BHLS02}).

Although calculating the precise capacity region may be difficult, a
more modest goal is to improve upon the inner and outer bounds given
in this paper.  In particular, known classical bounds that improve
upon Shannon's bounds \cite{Due79,Sch82,Sch83,Han84,ZBS86,HW89} might be
useful for finding improved quantum bounds.  Also, it would be
interesting to find conditions under which the inner and outer bounds
coincide.  Shannon showed that his inner and outer bounds coincide for
certain kinds of symmetric two-way classical channels \cite{Sha61},
so it is plausible to suppose that a similar result might hold in the
quantum case.

We have primarily considered the case of unlimited entanglement
assistance, but it would be interesting to consider unassisted
communication as well as the general case of finite entanglement
assistance.  Recently, Shor has given a protocol for classical
communication through a one-way quantum channel with limited
entanglement assistance that interpolates between the HSW capacity and
the entanglement-assisted capacity \cite{Shor04}.  
In addition, Harrow has obtained an expression for the one-way
classical communication capacity with finite entanglement assistance
for unitary two-way channels \cite{H03}.
It would be interesting to generalize these results to arbitrary
two-way channels.

Another approach is to consider particular families of channels to see
whether the capacity region is simpler for those channels.  
One such family is the set of two-way entanglement-breaking channels.
There are several possible definitions of a two-way
entanglement-breaking channel, but perhaps the simplest is that the
output state should be triseparable between Alice's output, Bob's
output, and any ancillas.  Unfortunately, it is not even clear how to
characterize such channels (as can be done for one-way
entanglement-breaking channels \cite{Rus03,HSR03}).
Some results have been obtained for other families of two-way quantum
channels, such as unitary two-way channels \cite{BHLS02,H03,HL04} and
feedback channels \cite{igor}.
Another special class of two-way channels, those that simply
distribute bipartite states, have been much better understood
\cite{family}.  

One way to obtain a better understanding of channel capacities is to
consider the problem of simulating a channel using a certain amount of
communication (in each direction) and entanglement.  Such {\em reverse
theorems} have been studied for one-way classical \cite{BSST01} and
quantum \cite{qrst} channels, and more recently for feedback channels
with restricted input sources \cite{igor}.  
In particular, reverse theorems can be useful for establishing bounds
on capacities \cite{BHLS02}. 
However, simple reverse theorems for general two-way channels seem
unlikely to exist.  For example, the communication costs (in each
direction) of any simulation must exceed the corresponding one-way
capacities, since the simulated channel can be used to achieve the
one-way capacity in either direction.  
As another example, the set of causal operations that are not
localizable \cite{BGNP01} cannot produce any communication, but any
such operation requires communication in at least one direction to
simulate, even with entanglement assistance. 

Finally, note that we have completely avoided the problems of
communicating quantum information through a two-way quantum channel
and of multi-way communication through $k$-partite quantum operations
with $k>2$.  These problems present further challenges for
understanding the the capabilities of quantum communication channels.

\section*{Acknowledgments}

We thank Aram Harrow for many helpful discussions, especially
regarding the protocol for achieving the entanglement-assisted one-way
capacity of a two-way channel.
Thanks also to Charles Bennett, whose insights on
entanglement-assisted communication via entanglement breaking channels
and the simulation of nonlocal boxes saved us from wandering in
unfruitful directions.
This work was initiated while AMC and DWL were at the IBM T.~J.\
Watson Research Center and HKL was at Magiq Technologies.  Part of
this work was done while AMC was at the MIT Center for Theoretical
Physics and while DWL was a visiting researcher at MSRI.  AMC was
supported in part by the Fannie and John Hertz Foundation, the
Cambridge--MIT Foundation, the DOE under cooperative research
agreement DE-FC02-94ER40818, and the NSA and ARDA under ARO contract
DAAD19-01-1-0656.  AMC and DWL received support from the NSF under
Grant No.\ EIA-0086038.  DWL received support from the Tolman
Foundation and the Croucher Foundation.  HKL received support from
NSERC, the CRC Program, CFI, OIT, PREA, and CIPI.


\end{document}